\documentclass{mn2e}
\usepackage{times}

\begin{document}
\journal{astro-ph/0401198}

\title[How many cosmological parameters?]
{How many cosmological parameters?}
\author[Andrew R. Liddle]
{Andrew R. Liddle\\ 
Astronomy Centre, University of Sussex, Brighton BN1 9QH, United Kingdom}
\maketitle
\begin{abstract}
Constraints on cosmological parameters depend on the set of 
parameters chosen to define the model which is compared with observational data.
I use the Akaike and Bayesian information criteria to carry out cosmological 
model selection, in order to determine the parameter set providing the preferred 
fit to the data. Applying the information criteria to the current cosmological 
data sets indicates, for example, that spatially-flat models are statistically 
preferred to closed models, and that possible running of the spectral index has 
lower significance than inferred from its confidence limits. I also discuss some 
problems of statistical assessment arising from there being a large number of 
`candidate' cosmological parameters that can be investigated for possible 
cosmological implications, and argue that 95\% confidence is too low a threshold 
to robustly identify the need for new parameters in model fitting. The best 
present description of cosmological data uses a scale-invariant ($n=1$) spectrum 
of gaussian adiabatic perturbations in a spatially-flat Universe, with the 
cosmological model requiring only five fundamental parameters to fully specify 
it. 
\end{abstract}
%%%%%%%%%%%%%%%%%%%%%%%%%%%%%%%%%%%%%%%%%%%%%%%%%%%%%%%%%%%%%%%%%%%%%%

\section{Introduction}

Since the release of microwave anisotropy data from the Wilkinson Microwave 
Anisotropy Probe (WMAP, Bennett et al.~2003), it has been widely acknowledged 
that cosmology has entered a precision era, with many of the key cosmological 
parameters being determined at the ten percent level or better. By now, a wide 
range of analyses have been published, uniting this dataset with other 
cosmological datasets such as galaxy power spectrum information from the Two 
degree field (2dF) survey or the Sloan Digital Sky Survey (SDSS).

While the various analyses are in broad agreement with one another, typically 
some 
differences do arise in the precise constraints, for two reasons. One is that 
separate analyses often use slightly different data compilations, which of 
course should lead to differing results, hopefully consistent within the 
uncertainties. However, further differences arise due to the choice of 
cosmological model made, usually meaning the number of cosmological parameters 
allowed to vary. The standard approach thus far has been to first choose the set 
of parameters to be varied on a fairly {\it ad hoc} basis, and then use a 
likelihood method to find the 
best-fit model and confidence ranges for those parameters. Some papers analyze 
several combinations of parameters, primarily with the aim of investigating 
how the parameter confidence ranges are affected by modifying these assumptions.

So far, however, there have been few attempts to allow the data to determine 
which combination of parameters gives the preferred fit to the data. This is the 
statistical problem of {\em model selection}, which arises across many branches 
of science; for example, in studies of medical pathologies, one wishes to know 
which set of indicators, out of many potential factors, are best suited to 
predicting 
patient susceptibility. The emphasis is usually on ensuring the elimination of 
parameters which play an insufficient role in improving the fit to the data 
available. A key tool is this area is {\em information criteria}, specifically 
the Akaike information criterion (Akaike 1974) and the Bayesian information 
criterion (Schwarz 1978). These have led to 
considerable advances in understanding of statistical inference and its relation 
to information theory; Akaike's 1974 paper now has over 3000 citations and is 
the subject of a complete textbook (Sakamoto, Ishiguro \& Kitagawa 1986). 
However, so far they seem to have had minimal application in astronomy --- 
keyword search on the abstracts of the entire astro-ph archive yields only four 
journal
papers (Mukherjee et al.~1998; Takeuchi 2000; Connolly et al.~2000; Nakamichi \& 
Morikawa 2003). 
In 
this paper I will apply the information criteria to the problem of selection of 
cosmological parameters.

\section{The information criteria}

The information criteria have a deep underpinning in the theory of statistical 
inference, but fortunately have a very simple expression. The key aim is to make 
an objective comparison of different models (here interpretted as different 
selections of cosmological parameters to vary) which may feature different 
numbers of parameters. Usually in cosmology a basic selection of `essential' 
parameters is considered, to which additional parameters might be added to make 
a more general model. It is assumed that the models will be compared 
to a fixed dataset using a likelihood method.

Typically, the introduction of extra parameters will allow an improved fit to 
the dataset, regardless of whether or not those new parameters are actually 
relevant.\footnote{In cosmology, a new parameter will usually be a quantity set 
to zero in the simpler base model, and as the likelihood is a continuous 
function of the parameters, it will increase as the parameter varies in either 
the positive or negative direction. However some parameters are restricted to 
positive values (e.g.~the amplitude of tensor perturbations), and in that case 
it may be that the new parameter does not improve the maximum likelihood.} A 
simple comparison of the maximum likelihood of different models will therefore 
always favour the model with the most parameters. The information criteria 
compensate for this by penalizing models which have more parameters, offsetting 
any improvement in the maximum likelihood that the extra parameters might allow.

The simplest procedure to compare models is the likelihood ratio test (Kendall 
\& Stuart 1979, ch. 24), which can 
be applied when the simple model is nested within a more complex model. The 
quantity $2\ln{\cal L}_{{\rm simple}}/{\cal L}_{{\rm complex}}$, where ${\cal 
L}$ is the maximum likelihood 
of the model under consideration, is approximately chi-squared distributed and 
standard statistical tables can be used to look up the significance of any 
increase in likelihood against the number of extra parameters introduced. 
However the assumptions underlying the test are often violated in astrophysical 
situations (Protassov et al.~2002).
Further, one is commonly interested in comparing models which are not nested.

The Akaike information criterion (AIC) is defined as
\begin{equation}
\mathrm{AIC} = -2\ln {\cal L} + 2k\,,
\end{equation}
where ${\cal L}$ is the maximum likelihood and 
$k$ the number of parameters of the model (Akaike 1974). The best model is 
the model which minimizes the AIC, and there is no requirement for the models to 
be nested. Typically, models with too few parameters 
give a poor fit to the data and hence have a low log-likelihood, while those 
with too many are penalized by the second term. The form of the AIC
comes from minimizing the Kullback--Leibler information entropy, which 
measures the 
difference between the true distribution and the model distribution. The AIC 
arises from an approximate minimization of this entropy; an explanation geared 
to astronomers can be found in Takeuchi (2000), while the full statistical 
justification can be found in Sakamoto et al.~(1986) and Burnham \& Anderson 
(2002).

The Bayesian information criterion (BIC) was introduced by Schwarz (1978), and 
can be defined as
\begin{equation}
\mathrm{BIC} = -2\ln{\cal L} + k \ln N \,,
\end{equation}
where $N$ is the number of datapoints used in the fit (in current cosmological 
applications, this will be of order one thousand). It comes from approximating 
the Bayes factor (Jeffreys 1961; Kass \& Raftery 1995), which gives the 
posterior odds of one model against another 
presuming that the models are equally favoured prior to the data fitting. 
Although expressed in terms of the maximum likelihood, it is therefore related 
to the integrated likelihood.

It is unfortunate that there are different information criteria in the
literature, which forces one to ask which is better.  Extensive Monte Carlo
testing has indicated that the AIC tends to favour models which have more
parameters than the true model (see e.g.~Harvey 1993; Kass \& Raftery 1995).
Formally, this was recognized in a proof that the AIC is `dimensionally
inconsistent' (Kashyap 1980), meaning that even as the size of the dataset tends
to infinity, the probability of the AIC incorrectly picking an overparametrized
model does not tend to zero.  By contrast, the BIC is dimensionally consistent,
as the second term in its definition ever more harshly penalizes
overparametrized models as the dataset increases in size, and hence the BIC does
always pick the correct model for large datasets.  Burnham \& Anderson (2002)
generally favour the AIC, but note that the BIC is well justified whenever the
complexity of the true model does not increase with the size of the dataset and
provided that the true model can be expected to be amongst the models
considered, which one can hope is the case in cosmology.  Accordingly, it seems
that that BIC should ordinarily be preferred.  Note though that for any likely
dataset $\ln N > 2$, and hence the AIC is always more generous towards extra
parameters than the BIC.  Hence the AIC remains useful as it gives an upper
limit to the number of parameters which should be included.

In either case, the absolute value of the criterion is not of interest, only the 
relative value between different models. A difference of 2 for the BIC is 
regarded as positive evidence, and of 6 or more as strong evidence, against the 
model with the larger value (Jeffreys 1961; Mukherjee et al.~1998).

\begin{table}
\caption{\label{t:base} Base parameters: those that appear essential for a 
successful cosmological model. Those 
below the line are in principle determinable from those above, but with present 
understanding are treated as free phenomenological parameters. Models based on 
these parameters alone provide an adequate fit to present cosmological data.}
\begin{tabular}{ll}
$\Omega_{{\rm m}}$ & matter density\\
$\Omega_{{\rm b}}$ & baryon density\\
$\Omega_{{\rm r}}$ & radiation density\\
$h$ & hubble parameter\\
$A$ & adiabatic density perturbation amplitude\\
\hline
$\tau$ & reionization optical depth\\
$b$ & bias parameter (or parameters)\\
\end{tabular}
\end{table}

\begin{table*}
\begin{minipage}{14cm}
\caption{\label{t:cand} Candidate parameters: those which might be relevant for 
cosmological 
observations, but for which there is presently no convincing evidence requiring 
them. They are listed so as to take the value zero in the base cosmological 
model. Those above the line are parameters of the background homogeneous 
cosmology, and those below describe the perturbations. Of the latter set, the 
first six refer to adiabatic perturbations, the next three to tensor 
perturbations, and the remainder to isocurvature perturbations.}
\begin{tabular}{ll}
$\Omega_k$ & spatial curvature\\
$N_\nu -3.04$ & effective number of neutrino species ({\sc cmbfast} 
definition)\\
$m_{\nu_i}$ & neutrino mass for species `$i$'\\
& [or more complex neutrino properties]\\
$m_{{\rm dm}}$ & (warm) dark matter mass\\
$w+1$ & dark energy equation of state\\
$dw/dz$ & redshift dependence of $w$\\
& [or more complex parametrization of dark energy evolution]\\
$c_{{\rm S}}^2-1$ & effects of dark energy sound speed\\
$1/r_{{\rm top}}$ & topological identification scale\\
& [or more complex parametrization of non-trivial topology]\\
$d\alpha/dz$ & redshift dependence of the fine structure constant\\
$dG/dz$ & redshift dependence of the gravitational constant\\
\hline
$n-1$ & scalar spectral index\\
$dn/d\ln k$ & running of the scalar spectral index\\
$k_{{\rm cut}}$ & large-scale cut-off in the spectrum\\
$A_{{\rm feature}}$ & amplitude of spectral feature (peak, dip or step) ...\\
$k_{{\rm feature}}$ & ... and its scale\\
& [or adiabatic power spectrum amplitude parametrized in $N$ bins]\\
$f_{{\rm NL}}$ & quadratic contribution to primordial non-gaussianity\\
 & [or more complex parametrization of non-gaussianity]\\
$r$ & tensor-to-scalar ratio\\
$r+8n_{{\rm T}}$ & violation of the inflationary consistency equation\\
$dn_{{\rm T}}/d\ln k$ & running of the tensor spectral index\\
${\cal P}_S$ & CDM isocurvature perturbation ...\\
$n_S$ & ... and its spectral index ...\\
${\cal P}_{S{\cal R}}$ & ... and its correlation with adiabatic perturbations 
...\\
$n_{S{\cal R}}-n_S$ & ... and the spectral index of that correlation\\
& [or more complicated multi-component isocurvature perturbation]\\
$G\mu$ & cosmic string component of perturbations\\
\end{tabular}
\end{minipage}
\end{table*}

The rather limited literature on cosmological model selection has thus far not 
used the information criteria, but has instead used the more sophisticated idea 
of Bayesian evidence (see e.g.~Jaynes 2003; MacKay 2003). This compares the 
total posterior 
likelihoods of the models, obtained as a product of the Bayes factor and the 
prior relative likelihood. This requires an 
integral of the likelihood over the whole model parameter space, which may be 
lengthy to calculate, but avoids the approximations used in the information 
criteria and also permits the use of prior information if required. It has been 
used in a variety of cosmological contexts by Jaffe (1996), Drell, Loredo \& 
Wasserman (2000), John \& Narlikar (2002), Hobson, Bridle \& Lahav 
(2002), Slosar et al.~(2003), Saini, Weller \& Bridle (2004), and Niarchou, 
Jaffe \& Pogosian (2004). 

\section{Application to present cosmological data}

\subsection{Choice of parameters}

Most of the recent work on cosmological parameters has chosen a particular 
parameter set or sets, and investigated parameter constraints when faced with 
different observational datasets. However, the information criteria ask how well 
different models fit the same dataset. First we need to decide which models to 
consider.

A useful division of parameters is into those which are definitely needed to 
give a reliable fit to the data, which I will call the base parameter set, and 
those which have proved irrelevant, or of marginal significance, in fits to the 
present data. The base parameter set is actually extraordinarily small, and 
given in Table~\ref{t:base}. At present it seems that a scale-invariant spectrum 
of adiabatic gaussian density perturbations, requiring specification of just a 
single parameter (the amplitude), is enough to give a good fit to the data. The 
Universe can be taken as spatially-flat, with the dark matter, baryon, and 
radiation densities requiring to be specified as independent parameters. The 
base model includes a 
cosmological constant/dark energy, whose density is fixed by the spatial 
flatness condition. To complete the parameter set, we need the Hubble constant. 
Accordingly, a minimal description of the Universe requires just five 
fundamental parameters.\footnote{To be more precise, this base model assumes all 
the parameters to be listed in Table~\ref{t:cand} are zero. Analyses may 
use different parameter definitions equivalent to those given here, for instance 
using the physical densities $\Omega h^2$ in place of the density parameters.} 
Further, the radiation density $\Omega_{{\rm r}}$ 
is directly measured at high accuracy from the cosmic microwave background 
temperature and is not normally varied in fits to other data. 

In addition to these fundamental parameters, comparisons with microwave 
anisotropy and galaxy power spectrum data require knowledge of the reionization 
optical depth $\tau$ and the galaxy bias parameter $b$ respectively. These are 
not fundamental parameters, as they are in principle computable from the above, 
but present understanding does not allow an accurate first-principles derivation 
and 
instead typically they are taken as additional phenomenological parameters to be 
fit from the data.  

Complementary to this base parameter set is what I will call the list of 
candidate parameters. These are parameters which are not convincingly measured 
with present data, but some of which might be required by future data. Many of 
them are available in model prediction codes such as {\sc cmbfast} (Seljak \& 
Zaldarriaga 1996).
Cosmological observations seek to improve the measurement of the base 
parameters, and also to investigate whether better data requires the promotion 
of any parameters from the candidate set into the standard cosmological model. 
Table~\ref{t:cand} shows a list of parameters which have already been discussed 
in the literature, and although already rather long is likely to be incomplete.

The upper portion of Table~\ref{t:cand} lists possible additional parameters 
associated with the background space-time, while the lower part contains those 
specifying the initial perturbations. The base cosmological model assumes these 
are all zero (as defined in the table), and indeed it is a perfectly plausible 
cosmological model that they are indeed all zero, with the sole exception of the 
neutrino masses, for which there is good non-cosmological evidence that they are 
non-zero. One should be fairly optimistic about learning something about 
neutrino masses from cosmology, which is why they are included as cosmological 
parameters. It is also possible that one day they might be pinned down 
accurately enough by other measurements that cosmologists no longer need to 
worry about varying them, and then neutrino masses will not be cosmological 
parameters any more than the electron or proton mass are.

It is of course highly unlikely that {\em all} the parameters on the candidate 
list will be relevant (if they were, observational data would have little 
chance of constraining anything), and on theoretical grounds some are thought 
much more likely than others. In most cases parameters can be added individually 
to the base model, but there are some dependences; for example, it doesn't make 
much sense to include spectral index running as a parameter unless the spectral 
index itself is included. Quite a lot of the parameters in Table~\ref{t:cand} 
have now been added to a base parameter set (usually not the one I have adopted 
here, however) and compared to observational data. There is also the 
possibility that the simultaneous inclusion of two extra parameters, which are 
unrelated, might significantly improve the fit where neither parameter 
separately did. This is hard to fully test as there are so many possible 
combinations. 

\subsection{Application to WMAP+SDSS data}

I will use the results from comparison of models to WMAP plus SDSS data given in 
Tegmark et al.~(2004, henceforth T04). Much of the analysis in that paper 
focusses on a simple parameter set called the `vanilla' model or sometimes the 
`six parameter' model. Confusingly, it actually features seven parameters (they 
do not count the bias parameter, although it is an independent fit parameter). 
They are not quite the set given in Table~\ref{t:base}; the radiation 
density parameter is omitted for reasons I explained above, while the spectral 
index $n$ is included as an independent parameter. However $n-1$ is not actually 
detected to be non-zero; its 1-sigma confidence range (table 4, column 6 of T04) 
is $0.952<n<1.016$. In light of the above discussion, we might expect that the 
information criteria reject the inclusion of $n-1$ as a useful parameter, and 
indeed that is the case. 

The $\chi^2$ values quoted by T04 are derived using the 
WMAP likelihood code [see Verde et 
al.~(2003) and Spergel et al.~(2003) for details] combined with a calculation of 
the likelihood from the SDSS data, and are defined as $-2\ln {\cal L}$. The 
total number of 
datapoints $N$ ({\em not} corrected for the number of parameters in the fit) is 
\mbox{$N=1367$} (899 WMAP temperature spectrum, 449 WMAP polarization 
cross-correlation, 19 SDSS). I note that their Markov chains were designed to 
estimate confidence intervals rather than to accurately determine the precise 
maximum likelihood, and a modest bias might occur from the maximum being less 
well pinpointed the greater the model dimensionality. Once the approximate 
locations of the maxima are determined via a Markov Chain Monte Carlo procedure, 
a variation on that method could be used to determine the maximum likelihoods 
accurately as an additional part of the data analysis process.

As seen in the upper two rows of Table~\ref{t:IC}, both information criteria 
prefer the base model, with $n$ fixed at one, as opposed to letting $n$ vary. As 
has been remarked before, there is presently no evidence that the parameter 
$n-1$ is needed to fit present data. T04 draw the same conclusion on subjective 
grounds, and refer to the base model as `vanilla lite'.

A similar argument applies to other cosmological parameters. Unfortunately the 
other models analyzed by T04 include variation of $n$ (their table 3) and so 
other parameters are not directly compared with the base model, but anyway the 
trend seen in Table~\ref{t:IC} is clear --- the more parameters included the 
higher the AIC and BIC as compared to the base model.  The need for these 
additional parameters is strongly rejected by the information criteria, 
particularly the BIC which strongly penalizes additional parameters for a 
dataset of this size.\footnote{It is interesting to note that recent 
applications of the Bayesian evidence to cosmological model selection have also 
found no significant evidence against the simplest model considered (Slosar et 
al.~2003; Saini et 
al.~2004; Niarchou et al.~2004).} For example, the information criteria reject 
the need for 
$\Omega_k$ as an independent parameter, instead identifying spatially-flat 
models as the preferred description of the data.

\begin{table}
\caption{\label{t:IC} AIC and BIC for the various models, with likelihood values 
taken from tables 3 and 4 of T04. The upper two rows compare the base model with 
the addition of $n$ as an extra parameter. The lower entries show various other 
combinations of parameters. I drop the radiation density from the parameter list 
as it is not needed to fit these data.}
\begin{tabular}{lcccc}
Model & parameters & $-2\ln {\cal L}$ & AIC & BIC\\
\hline
Base model & 6 & 1447.9 & 1459.9 & 1491.2\\
Base + $n$ & 7 & 1447.2 & 1461.2 &1497.7 \\
\hline
Base + $n$,$\Omega_k$ & 8 & 1445.4 & 1461.4 &1503.2 \\
Base + $n$,$r$ & 8 &1446.9 & 1462.9 & 1504.7\\
Base + $n$,$r$,$\frac{dn}{d\ln k}$,$\Omega_k$ & 10 & 1444.4 & 1464.4 & 1516.6 \\
\end{tabular}
\end{table}

\section{Candidate parameters and statistical significance}

The information criteria are clearly a powerful tool for establishing the 
appropriate set of cosmological parameters. How do they relate to the standard 
approach in cosmology of looking at confidence levels of parameter detection?

Use of fairly low confidence levels, such as 95\%, to identify new parameters is 
inherently very risky because of the large number of candidate parameters. If 
there were only one candidate parameter and it were detected at 95\% confidence, 
that certainly be interesting. However there are many possible parameters, and 
if one analyzes a several of them and finds one at 95\% confidence, then one can 
no longer say that the base model is ruled out at that level, {\em because there 
were several different parameters any of which might, by chance, have been at 
its 95\% limit}. As an extreme example, if one considered 20 parameters it would 
be no surprise at all to find one at 95\% confidence level, and that certainly 
wouldn't mean the base model was excluded at that confidence. Consequently the 
true statistical significance of a parameter detection is always likely to be 
less than indicated by its confidence levels (e.g.~Bromley \& Tegmark 2000). 
This issue can arise 
both within a single paper which explores many parameters, and in a broader 
sense because the community as a whole investigates many different parameters.

This is a form of publication bias --- the tendency for authors to 
preferentially submit, and editors to preferentially accept, papers showing 
positive statistical evidence. This bias is well recognized in the field of 
medical trials (see e.g.~Sterne, Gavaghan \& Egger 2000), where it can literally 
be a matter of life and death and tends to lead to the introduction of 
treatments which are at best ineffectual and may even be harmful. The stakes are 
not so high in cosmology, but one should be aware of its possible effects. 
Publication bias comes in several forms, for example if a single paper analyzes 
several parameters, but then focusses attention on the most discrepant, that in 
itself is a form of bias. The more subtle form is where many different 
researchers examine different parameters for a possible effect, but only those 
who, by chance, found a significant effect for their parameter, decided to 
publicize it strongly.

Publication bias is notoriously difficult to allow for, as it mainly arises due 
to unpublished analyses of null results. However a useful guide comes from 
considering the number of parameters which have been under discussion in the 
literature. Given the list in Table~\ref{t:cand}, it is clear that, even if the 
base cosmological model is correct, there are enough parameters to be 
investigated that one should not be surprised to find one or two at the 
95\% confidence level. 

I conclude that when considering whether a new parameter should be transferred 
from the candidate parameter list to the base parameter list, a 95\% confidence 
detection should not be taken as persuasive evidence that the new parameter is 
needed. Because there are so many candidate parameters, a more powerful 
threshold is needed. The BIC provides a suitably stringent criterion, whereas 
this line of argument supports the view that the AIC is too weak a criterion for 
cosmological model selection.

Another subtle point relating to cosmological data is the inability to 
fully repeat an experiment. Conventionally in statistics, once a dataset 
has identified an effect which looks interesting (e.g.~spectral index running at 
95\% confidence), one is expected to throw away all that data and seek 
confirmation from a completely new 
dataset. This procedure is necessary to minimize publication bias effects, and 
failure to follow it is regarded as poor practice. Unfortunately, for the 
microwave anisotropies 
much of the noise comes from cosmic variance rather than instrumental effects, 
and so remeasuring does not give an independent realization of statistical 
noise. For example, if one analyzes the second-year WMAP data (once it becomes
available) separately from the first-year data, there will be a tendency for the 
same cosmological parameter values to be obtained. Finding the same outlying 
parameter values therefore will have less statistical significance than were the 
datasets genuinely independent. Even Planck data will have noise significantly 
correlated to WMAP data in this sense, and properly allowing for that in 
determining statistical significance of parameter detections would be tricky. 
This supports the use of information criteria for model selection, rather than 
parameter confidence levels.

\section{Conclusions}

Various conclusions can be drawn from the information criterion approach. Most 
importantly, they provide a simple objective criterion for the inclusion of new 
parameters into the standard cosmological model. For example, it is sometimes 
said that the WMAP analysis actually mildly favours a closed cosmological model, 
as their best-fit value is $\Omega_k = 1.02 \pm 0.02$ (at 1-sigma). However, the 
information criteria lead to the opposite conclusion: they say that the most 
appropriate conclusion to draw is that the spatial curvature is not needed as a 
parameter, and hence it is more likely that the observations were generated in a 
spatially-flat Universe. That's not to say that future observations might not 
show that the Universe is closed, but a much higher significance level 
than 1-sigma is needed before it becomes the best description of the data in 
hand. Similar 
arguments can be applied also to parameters such as running of the spectral 
index; even in the absence of controversy over the use of lyman-alpha forest 
data, it seems likely that the information criteria would reject the 
running as a useful parameter (I can't test it, as the WMAP team were unable to 
quote a maximum likelihood due to unknown error covariances). In general, a 95\% 
`detection' of a particular new parameter cannot be taken to imply that the base 
model, without that parameter, is ruled out at anything like that significance.

According to the information criteria, the best current cosmological model 
features only five fundamental parameters and two phenomenological ones, as 
listed 
in Table~\ref{t:base}. While there is an elegant simplicity to this model which 
is satisfying, such simplicity does come at a cost, because the cosmological 
parameters are what tells about the physical processes relevant to the evolution 
of the Universe. That there are so few parameters is telling us that there is 
very little physics that we are currently able to probe observationally. 
Accordingly, we 
should be hoping that new observational data is powerful enough to 
promote parameters from the candidate list to the base list; for example, we 
won't be able to say anything quantitative about how cosmological inflation 
might have taken 
place unless $n-1$, and ideally $r$ as well, make their way into the standard 
cosmological model.

The information criteria appear well suited to providing an objective criterion 
for the incorporation of new parameters, and have had considerable testing 
across many scientific disciplines. The BIC appears to be preferred to the AIC 
for cosmological applications. For the size of the current dataset the 
BIC penalizes extra parameters very strongly, indicating that a very 
high-significance 
detection is needed to justify adoption of a new parameter.

\section*{Acknowledgments}
This research was supported in part by PPARC. I thank Charles Goldie for 
directing me to the literature on the information criteria, and thank Sarah 
Bridle,  
Martin Kunz, Sam Leach, Max Tegmark, and the referee H{\aa}vard Sandvik for 
helpful discussions and comments.

%\bsp


\begin{thebibliography}{}
\bibitem[Akaike 1974]{Akaike} Akaike H., 1974, IEEE Trans. Auto. Control, 
	19, 716
\bibitem[Bennett et al.~2003]{Beetal03} Bennett C. L. et al.~(the WMAP Team), 
	2003, ApJS, 148, 1
\bibitem[Bromley \& Tegmark 2000]{BT00} Bromley B. C., Tegmark M., 2000, ApJL,
	524, L79
\bibitem[Burnham \& Anderson 2002]{BA02} Burnham K. P., Anderson D. R., 2002,
	{\em Model selection and multimodel inference}, 2nd ed.,
	Springer-Verlag, New York
\bibitem[Connolly et al.~2000]{con00} Connolly A. J., Genovese C., Moore A. W.,
	Nichol R. C., Schneider J., Wasserman L., 2000, astro-ph/0008187
\bibitem[Drell. et al.~2000]{Dre00} Drell P. S., Loredo T. J., Wasserman I.,
	2000, ApJ, 530, 593
\bibitem[Harvey 1993]{Harvey} Harvey A. C., 1993, {\em Time series models},
	2nd ed., Prentice Hall, Hertfordshire (UK)
\bibitem[Hobson et al.~2002]{Hob02} Hobson M. P., Bridle S. L., Lahav O., 2002,
	MNRAS, 335, 377
\bibitem[Jaffe 1996]{Jaf96} Jaffe A., 1996, ApJ, 471, 24
\bibitem[Jaynes 2003]{Jay03} Jaynes E. T., 2003, {\em Probability theory: 
	the logic of science}, Cambridge University Press
\bibitem[Jeffreys 1961]{Jeff} Jeffreys H., 1961, {\em Theory of probability},
	3rd ed., Oxford University Press
\bibitem[John \& Narlikar]{JN} John M. V., Narlikar J. V., 2002, Phys. Rev.
	D, 65, 043506
\bibitem[Kashyap 1980]{Kashyap} Kashyap R., 1980, IEEE Trans. Auto. Control, 25,
	996
\bibitem[Kass \& Raftery 1995]{KR95} Kass R. E., Raftery A. E., 1995, Journ.
	American Stat. Assoc., 90, 773
\bibitem[Kendall \& Stewart 1979]{KS79} Kendall M., Stuart A., 1979, {\em
	Advanced theory of statistics}, vol.~2, 4th ed., Griffin, London
\bibitem[MacKay 2003]{mackay} MacKay D. J. C., 2003, {\em Information theory,
	inference, and learning algorithms}, Cambridge University Press
\bibitem[Mukherjee et al.~1998]{Muk98} Mukherjee S., Feigelson E. D., 
	Babu G. J., Murtagh F., Fraley C., Raftery A., 1998, ApJ, 508, 314
\bibitem[Nakamichi and Morikawa 2003]{Nak03} Nakamichi A., Morikawa M., 2003,
	astro-ph/0304301
\bibitem[Niarchou et al.~2004]{Nia04} Niarchou A., Jaffe A. H., Pogosian L.,
	2003, Phys. Rev. D, 69, 063515
\bibitem[Protassov et al.~2002]{Pro} Protassov R., van Dyk D. A., Connors A.,
	Kashyap V. L., Siemiginowska A., 2002, ApJ, 571, 545
\bibitem[Saini et al.~2004]{Sai04} Saini T. D., Weller J., Bridle S. L., 2004,
	MNRAS, 348, 603
\bibitem[Sakamoto et al.~1986]{Saketal} Sakamoto Y., Ishiguro M., Kitagawa G.,
	1986, {\em Akaike information criterion statistics}, Kluwer academic 
	publishers, Dordrecht
\bibitem[Schwarz 1978]{Schwarz} Schwarz G., 1978, Annals of Statistics, 
	5, 461
\bibitem[Seljak \& Zaldarriaga 1996]{SZ96} Seljak U., Zaldarriaga M., 1996, ApJ,
	469, 1
\bibitem[Slosar et al.~2003]{Slo2003} Slosar A. et al., 2003, MNRAS, 341, L29
\bibitem[Spergel et al.~2003]{Sper} Spergel D. N. et al.~(the WMAP Team), 2003, 
	ApJS, 148, 175
\bibitem[Sterne et al.~2003]{Ster} Sterne J. A. C., Gavaghan D., Egger M.,
	2000, Journal of Clinical Epidemiology, {\bf 53}, 1119
\bibitem[Tegmark et al.~2004]{Teg04} Tegmark M. et al.~(the SDSS Collaboration), 
	2004, Phys. Rev. D, 69, 103501 (T04)
\bibitem[Takeuchi 2000]{Tak} Takeuchi T. T., 2000, Astrophys. Space Sci., 271,
	213
\bibitem[Verde et al.~2003]{Veetal03} Verde L. et al., 2003, ApJS, 148, 195

\end{thebibliography}
\end{document}